\documentclass[11pt]{article}
\usepackage{graphicx}  % standard LaTeX graphics tool
\usepackage{amssymb,latexsym}%,amsmath}
\usepackage[mathscr]{eucal}
\usepackage{citesort}
\usepackage{url}
\newcommand{\zgg} {\mathring{g}} % koneksja metryki tla
\newcommand{\zh} {\mathring{h}} % koneksja metryki tla
\newcommand{\mcN}{\mycal N}
\newcommand{\FS}       %{F_1} %
                  {F}
\newcommand{\HS} %{F_2}
       {H_{\mbox{\scriptsize volume}}}

{\ptc{this should be removed in the oberwolfach version}}%

\newcommand{\zDelta}{\Delta_{\mathring{h}}}%
\newcommand{\zD}{\mathring{D}}%
\newcommand{\zA}{\mathring{A}}%
\newcommand{\eeal}[1]{\label{#1}\end{eqnarray}}
\newcommand{\bed}{\begin{deqarr}}
\newcommand{\eed}{\end{deqarr}}
\newcommand{\bedl}[1]{\begin{deqarr}\label{#1}}
\newcommand{\eedl}[2]{\arrlabel{#1}\label{#2}\end{deqarr}}

\newcommand{\mcO}{{\mycal O}}

\newcommand{\bel}[1]{\begin{equation}\label{#1}}
\newcommand{\bea}{\begin{eqnarray}}
\newcommand{\bean}{\begin{eqnarray}\nonumber}
\newcommand{\beal}[1]{\begin{eqnarray}\label{#1}}
\newcommand{\eea}{\end{eqnarray}}
\newcommand{\Eq}[1]{Equation~\eq{#1}}

\newcommand{\be}{\begin{equation}}
\newcommand{\eeq}{\end{equation}}
\newcommand{\ee}{\end{equation}}
\newcommand{\beqa}{\begin{eqnarray}}
\newcommand{\eeqa}{\end{eqnarray}}
\newcommand{\beqan}{\begin{eqnarray*}}
\newcommand{\eeqan}{\end{eqnarray*}}
\newcommand{\ba}{\begin{array}}
\newcommand{\ea}{\end{array}}

\newcommand{\mcM}{{\mycal M}}

\DeclareFontFamily{OT1}{rsfs}{}
\DeclareFontShape{OT1}{rsfs}{m}{n}{ <-7> rsfs5 <7-10> rsfs7 <10-> rsfs10}{}
\DeclareMathAlphabet{\mycal}{OT1}{rsfs}{m}{n}
\newcounter{mnotecount}[section]

\renewcommand{\themnotecount}{\thesection.\arabic{mnotecount}}

\newcommand{\mnote}[1]{}%
\newcommand{\warn}[1]%{}
{\protect{\stepcounter{mnotecount}}$^{\mbox{\footnotesize
$%\!\!\!\!\!\!\,
\bullet$\themnotecount}}$ \marginpar{%\color{red}%
\raggedright\tiny\em
$\!\!\!\!\!\!\,\bullet$\themnotecount: {\bf Warning:} #1} }

\newcommand{\R}{\mathbb R}

\newcommand{\eq}[1]{(\ref{#1})}

\newcommand{\ptc}[1]{\mnote{{\bf ptc:}#1}}

\newcommand{\beqar}{\begin{deqarr}}
\newcommand{\eeqar}{\end{deqarr}}

\newcommand{\beaa}{\begin{eqnarray*}}
\newcommand{\eeaa}{\end{eqnarray*}}

\newcommand{\zg}{\mathring{g}}

\newcommand{\zR}{\mathring{R}}

\begin{document}

\title{On non-existence of static vacuum black holes with degenerate components of the
event horizon}

\author{Piotr T.\ Chru\'sciel\thanks{Partially supported by a Polish
Research Committee grant 2 P03B 073 24. E-mail
    \protect\url{Piotr.Chrusciel@lmpt.univ-tours.fr}, URL
    \protect\url{ www.phys.univ-tours.fr/}$\sim$\protect\url{piotr}}
  \\ LMPT,
F\'ed\'eration Denis Poisson\\
%Facult\'e des Sciences\\ Parc
%  de Grandmont\\ F37200
Tours
%, France
  \\
  \\
  Harvey S. Reall\thanks{{ E--mail}: \protect\url{Harvey.Reall@nottingham.ac.uk}}
  \\School of Physics and Astronomy, University of Nottingham\\Nottingham NG7 2RD, United Kingdom
\\
\\
  Paul Tod\thanks{{ E--mail}: \protect\url{paul.tod@st-johns.oxford.ac.uk}}
\\
Mathematical Institute and St John's College\\ Oxford}

\maketitle
\begin{abstract}
We present a simple proof of the non-existence of degenerate components of the event
horizon in static, vacuum, regular, four-dimensional black hole spacetimes. We discuss the generalization to higher dimensions and the inclusion of a cosmological constant.
\end{abstract}

%\section{Introduction}
%\label{introduction}

The classical proof of uniqueness of static vacuum black
holes~\cite{bunting:masood} assumes that all components of the event
horizon are non-degenerate. The argument has been extended\footnote{A
static configuration with all components degenerate is easy to exclude
using Komar integrals and the positive energy theorem,
see~\cite[Section~4]{Chstatic} for precise statements. However, one
also wants to exclude solutions with both degenerate and
non-degenerate components.}  to include degenerate
components~\cite{Chstatic} by studying the orbit-space geometry near
the event horizon, and applying an appropriate version of the positive
energy theorem~\cite{BartnikChrusciel1}. The object of this note is to
give an alternative proof of non-existence of degenerate components in
four dimensional space-times. Our approach is inspired by the analysis
of supersymmetric black holes in~\cite{Reall:2002bh}.

Consider, thus, a four-dimensional static vacuum black hole space-time
$(\mcM,g)$. We shall assume that the regularity hypotheses needed in
the \emph{black hole topology theorem} hold, the reader is referred to
points (1) and (2) of ~\cite[Theorem~3]{ChWald} for details. It
follows that each connected component of the horizon has spherical
topology.

We assume, for contradiction, that $(\mcM,g)$ contains a
\emph{degenerate} component $\mcN$ of the event
horizon. By \cite{CarterJMP} or \cite{Vishveshwara} the static Killing
vector field $X$ is tangent to the generators of $\mcN$.
Following~\cite{VinceJimcompactCauchyCMP}, we introduce Gaussian null
coordinates near $\mcN$, in which the metric takes the form
\bel{GNC1} g=r \varphi dv^2 + 2dv dr + 2r h_a dx^a dv + h_{ab}dx^a
dx^b\;.\ee
%The metric $h=h_{ab}dx^a dx^b$ will be referred to as
%\emph{the space metric on $\mcN$}.
(These coordinates can be introduced for any Killing horizon, not
necessarily static, in any number of dimensions).  The horizon is given by the equation
$\{r=0\}$. The Killing vector $X$ equals $\partial_v$, with norm
$$g(X,X) = r\varphi\;,$$ so that the surface gravity equals
$\kappa=-\partial_r (r\varphi)$.  The degeneracy condition is
$\kappa=0$, hence $\partial_r \varphi$ vanishes on $\mcN$. It follows
that
$$\varphi=Ar$$ for some function $A=A(r,x^a)$.

To simplify the calculations it is convenient to consider the
\emph{near horizon geometry} $\zg$, defined as follows \cite{Reall:2002bh}. Let
$\varepsilon>0$ and consider the family of metrics $g^\varepsilon$
defined by replacing $r$ by $\varepsilon r$ and $v$ by
$v/\varepsilon$ in \eq{GNC1}:
\bel{GNC2} g^\varepsilon =r^2 A^\varepsilon dv^2 + 2dv dr + 2r h^\varepsilon_a dx^a dv + h^\varepsilon_{ab}dx^a
dx^b\;.\ee with
$$A^\varepsilon=A(\varepsilon r,x^a)\;, \quad
\mbox{etc}\;.$$ Clearly the $g^\varepsilon$'s converge, as
$\varepsilon$ tends to zero, to a metric $\zg$ of the form
\beal{GNC4} & \zgg =r^2 \zA dv^2 + 2dv dr + 2r \zh_a dx^a dv + \zh_{ab}dx^a
dx^b\;, &\\
&\partial_r \zA=\partial_r \zh_a=\partial_r \zh_{ab}=\partial_v
\zA=\partial_v \zh_a=\partial_v \zh_{ab}=0
 \;.\eeal{GNC5} We have $\zh_{ab}=h_{ab}|_{r=0}$,
 $\zh_{a}=h_{a}|_{r=0}$, $\zA=A|_{r=0}$, so that $\zg$ encodes
 information about the values of $h_{ab}$, $h_{a}$ and $A$ at
 $\mcN$.

The vacuum Einstein equations imply (see~\cite[eq.~(2.9)]{VinceJimcompactCauchyCMP} in dimension four and \cite[eq.~(5.9)]{LP2} in
higher dimensions)
\bel{vEe} \zR_{ab} = \frac 12 \zh_{a}\zh_{b} -  \zD_{(a
}\zh_{b)}\;,\ee where $\zR_{ab}$ is the Ricci tensor of $\zh_{ab}$,
and $\zD$ is the covariant derivative thereof. They also determine
$\zA$ uniquely in terms of $\zh$ and $\zh_{ab}$ (this equations
follows again e.g. from~\cite[eq.~(2.9)]{VinceJimcompactCauchyCMP} in
dimension four, and can be checked by a calculation in all higher
dimensions):
\bel{Asol}
\zA = \frac{1}{2} \zh^{ab} \left( \zh_a \zh_b -  \zD_a \zh_b  \right).
\ee
We should note that, so far, our analysis applies
to any degenerate Killing horizon (not necessarily static) in four or
more dimensions. Equation \eq{vEe} also arises in the study of
vacuum degenerate isolated horizons~\cite{Ashtekar:2001jb,LP1,LP2,LJP}.

In the static case of interest here, note that staticity of $g$
implies staticity of $\zg$. If we let $$X^\flat:=\zg_{\mu\nu}X^\nu
dx^\mu = \zg_{\mu v} dx^\mu=r^2
\zA dv +  dr + r \zh_a dx^a\;,$$ then the staticity condition
$X^\flat\wedge dX^\flat=0$ leads to \beaa 0& =& (r^2 \zA dv +  dr + r
\zh_a dx^a)\wedge(d(r^2 \zA dv) +  d r\wedge(\zh_a   dx^a)+ r
d(\zh_a dx^a))
\\ &=& r dr \wedge d(\zh_a dx^a) +O(r^2)\;,\eeaa implying $d(\zh_a
dx^a)=0$. We now return to four dimensions.  Simple connectedness of
$S^2$ guarantees the existence of a function $\lambda$ such that
\bel{lambdef}\zh_a dx^a = d\lambda\;.\ee
\Eq{vEe} can thus be rewritten as
\bel{vEe2} \zR_{ab} = \frac 12 \zD_{a}\lambda \zD_{b}\lambda -  \zD_{a}\zD_{ b}\lambda\;.\ee
%Taking a trace and integrating over $N$ shows that
% \bel{vEe3} \int_N\zR = \frac 12 \int_N|d\lambda|^2\ge
%0\;.\ee
 Set $\psi = e^{-\lambda/2}$, then
\bel{in3} \psi \zR_{ab}=2\zD_{a}\zD_{ b}\psi\;,\ee and taking a
 trace gives
%$$\Delta_{\zh} \psi = 2 \zR \psi\ge 0\;,$$
%since $\psi$ is positive by definition, and $\zR$ is positive by
%\eq{vEe3}.
\bel{lapleq}2\Delta_{\zh} \psi =  \zR \psi\;.\ee In dimension two we have $\zR_{ab}=\zR
\zh_{ab}/2$; inserting this into \eq{in3}, applying $\zD^a$ to the
resulting equation, and commuting derivatives one
obtains\footnote{This is a special case of a result for stationary
degenerate horizons obtained in~\cite{LP1}.}
\bel{Rpsi3}
\zD_b(\zR
\psi^3)=0\;.
\ee
It follows that $\zR
\psi^3$ has constant sign or vanishes, so that $\zR$ has constant sign
or vanishes since $\psi$ is strictly positive by definition. On a
compact manifold this is compatible with \eq{lapleq} only if $\zR
\psi=0$ and $\psi $ is a constant, thus $\lambda$ is constant and
\eq{vEe2} shows that $\zh$ is flat.  This gives a contradiction, as
there are no flat metrics on $S^2$ by the Gauss-Bonnet theorem.
Hence, no degenerate components are possible, as claimed.

Recall that the Curzon-Scott-Szekeres~\cite{ScottSzekeresII} black
holes provide examples of vacuum static black holes with flat
degenerate horizons. However, those space-times are nakedly singular,
so that the topology theorem does not apply. We also note that while
the near horizon geometry of those black holes is compatible with our
result, this fact can \emph{not} be deduced from our analysis, as the
horizon there does not have compact cross-sections.

Some comments about higher dimensions are in order. First, the proof
in~\cite{Chstatic} generalises immediately to any space-time dimension
greater than or equal to four. On the other hand, the hypothesis of
space-time dimension four was essential in several steps of the
current argument so in higher dimensions we must proceed differently.
%\Eq{vEe} still holds~\cite[eq.~(5.9)]{LP2}.
Without the approach in~\cite{Chstatic}, there is no reason to expect the horizon to be simply connected, so
that a globally defined potential $\lambda$ might fail to exist. Assuming, first, that $\lambda$ in
\eq{lambdef} exists, one finds again
\eq{vEe2}. Taking a divergence of \eq{in3} and using the contracted
 Bianchi identity $\zD_{a}\zR^{a}{_b}= \zD_b
\zR/2$ one obtains
 \bel{VanD}\zD_a(|\zD\psi|^2+ \frac 12 \zR \psi^2)=0\;.\ee
Hence there exists a constant $C$ such that $$\zR=
\psi^{-2}(C-2|\zD\psi|^2)\;.$$ Inserting this into \eq{lapleq} one is
led to \bel{genL} \zDelta \psi^2 = C\;,\ee which is possible on a compact
manifold if and only if $C$ vanishes, $\psi$ is constant, and then
$\zh_{ab}$ is Ricci flat by \eq{in3}.

In general, let $\{\mcO_i\}_{i\in I}$ be an open cover by simply
connected sets, with associated potentials $\lambda_i$ and
$\psi_i$. Equation~\eq{VanD} shows that there exist constants $C_i$
such that on each $\mcO_i$ we can write
\bel{Vand2}\zR=
C_i\psi_i^{-2}-2|\zD\ln(\psi_i)|^2=
C_i\psi^{-2}_i-\frac 12\zh^{ab}\zh_a \zh_b \;.\ee
It follows that on each intersection $\mcO_i\cap \mcO_j$ we have
\bel{int}C_i\psi^{-2}_i=C_j\psi^{-2}_j\;,\ee
which implies that all the $C_i$'s have the same sign.

Suppose that
there exists $i_0$ such that $C_{i_0}\ne 0$, then $C_i\ne 0$ for all
$i$ by \eq{int}. The $\lambda_i$'s are defined up to the addition of a
constant, which implies that the $\psi_i$'s are defined up to a
multiplicative constant, and by rescaling we can obtain either $C_i=1$
for all $i$, or $C_i=-1$ for all $i$. It then follows from \eq{int} that $\psi_i = \psi_j$ on each intersection, i.e., $\psi$ is globally defined after all.
But then the previous argument shows $C=0$, hence $C_i=0$ for all $i$, a contradiction.

So in fact all the $C_i$'s  vanish and, by \eq{Vand2},
\bel{Vand3}\zR=-\frac 1 2\zh^{ab}\zh_a \zh_b\;.\ee
But the trace of \eq{vEe} gives $\zR=\zh^{ab}\zh_a \zh_b/2
+\zD^a\zh_a$, which upon integration on a cross-section of the horizon
gives
 $$\int \zR = \frac 12 \int \zh^{ab}\zh_a \zh_b \ge 0\;.$$ This
is compatible with \eq{Vand3} if and only if $\zh_a=0$.

Thus, we have shown, in all
space-time dimensions, that
\emph{static,  degenerate, vacuum Killing horizons with compact
spacelike sections have vanishing scalar $\zA$ and rotation form
$\zh_adx^a$ and are spatially Ricci flat}\footnote{Space-time
dimension three is covered by applying the four dimensional result to
$M\times S^1$ with the product metric.}.
The near-horizon geometry is the product of flat
space with a compact Ricci flat space. This is not known to lead to a
contradiction with asymptotic flatness except in space-time dimension
four (compare~\cite{GallowaySchoen}), unless one invokes the arguments
in~\cite{Chstatic}, which we wanted to sidestep to start with.

It turns out that one can derive the local form of the metric
$\zh_{ab}$ in space-time dimension four, when compactness of the
space-sections of the horizon is not assumed. In four dimensions, equations \eq{Rpsi3} and \eq{VanD} establish that, regardless of compactness, there exist constants
$\alpha$ and $\beta$ such that $$|\zD\psi|^2= \alpha
+\frac\beta\psi\;,\qquad
\Delta_{\zh}\psi=-\frac{\beta}{\psi^2}\;.$$ Assume $d\psi\not \equiv
0$, then on any open set on which $d\psi$ has no zeros the metric can
be written in the form $$\zh_{ab}dx^adx^b=
\frac{d\psi^2}{|\zD\psi|^2}+H(\varphi,\psi)d\varphi^2\;.$$ Calculating
$\Delta_{\zh}\psi$ one finds $H=\gamma(\varphi) |\zD\psi|^2$ for some function
$\gamma$. Redefining $\varphi$ one obtains, locally \bel{genmet} \zh_{ab}dx^adx^b=
\frac{d\psi^2}{\alpha +\frac\beta\psi}+(\alpha
+\frac\beta\psi)d\varphi^2\;.\ee It is straightforward now to check
that \eq{in3} holds for all $\alpha^2+\beta^2\ne 0$.

Now, the full near-horizon metric $\zg$ is closely
related to a generalised Schwarzschild solution. To see this, note that we can
use the freedom to rescale $\psi$ and $\phi$ to arrange $\alpha = k
\in \{1,0,-1\}$.  We shall also introduce the suggestive notation
$\beta = -2M$, and consider the following metric:
\bel{genSchw}
 ds^2 = -U(R) dt^2 + U(R)^{-1} dR^2 + R^2 d\Sigma_k^2,
\ee where $U(R) = k - 2M/R$ and $d\Sigma_k^2$ is a two-dimensional
Riemannian metric with Ricci scalar $2k$. Setting $R=\psi$ and
$t=i\varphi$, the first two terms in the metric reproduce the local
solution for $\zh_{ab}$ obtained above. The full near-horizon
geometry $\zg$ is obtained by a further analytic continuation in
which $d\Sigma_k^2$ becomes a {\it Lorentzian} metric of Ricci
scalar $2k$, i.e., two-dimensional de Sitter, Minkowski or anti-de
Sitter space-time for $k=1,0,-1$ respectively. The degenerate
horizon corresponds to a Killing horizon in this two-dimensional
space-time. Note that  $\zh_{ab}$ is singular except in the
trivial (flat\footnote{The corresponding space-time metric
\eq{GNC1} is also flat, with a singularity at $\psi=0$ which
can be gotten rid of by a coordinate transformation.}) 
case $k=1, M=0$ and the case $k=1,M>0$
in which it describes the familiar ``cigar" geometry of the
``Euclideanized'' Schwarzschild solution. 

Our analysis can be extended to include a cosmological constant
$\Lambda$. This produces additional terms $\Lambda \zh_{ab}$ and
$\Lambda$ on the right-hand-sides of \eq{vEe} and \eq{Asol}
respectively. Then
\begin{itemize}
\item For spatially compact horizons, in all space-time dimensions
greater than or equal to four, for negative $\Lambda$ we again obtain
the existence of a globally defined potential $\psi$. Applying a
maximum principle to the $\Lambda$-analogue of \eq{genL} one finds
that $\psi$ is constant and $\zR_{ab}=\Lambda \zh_{ab}$. The
near-horizon geometry $\zg$ is the product of two-dimensional anti-de
Sitter space with a compact Einstein space of negative curvature.
\item In space-time dimension four, whatever $\Lambda$,
\begin{itemize}
\item if $\psi$ is not constant,  we obtain 
\bel{Feq}
|\zD\psi|^2= \alpha +\frac\beta\psi-\frac \Lambda3\psi^2 =:F(\psi)
\ee
and, locally, 
\bel{genmet2} \zh_{ab}dx^adx^b=
\frac{d\psi^2}{\alpha +\frac\beta\psi-\frac \Lambda3
\psi^2}+\Big(\alpha +\frac\beta\psi-\frac \Lambda3
\psi^2\Big)d\varphi^2\;.\ee  
The near-horizon geometry $\zg$ is an
analytically continued version of the $\Lambda$-generalized
Schwarzschild solution (equation \eq{genSchw} with $U(R) = k -
2M/R-\Lambda R^2/3$).

Assuming compactness of the cross-section, the strictly positive
function $\psi$ has at least one maximum and at least one distinct
minimum so there exist $0<a<b$ such that $F(a)=F(b)=0$; for
$\Lambda>0$ this enforces $\beta<0$. Regularity at the zeros of $F$
(see~\cite[end of Section~2]{CT} for a careful treatment of a similar
problem, or the arguments around (3.6) in~\cite{GinspargPerry})
imposes
 $$F'(a)+F'(b)=0\;.$$ Elementary algebra leads to $a=b$, a
 contradiction. Therefore, in the case of positive $\Lambda$ as well,
 horizons with non-trivial $\psi$ and with compact cross-sections do
 not exist.

\item if instead  $\zh_a
\equiv 0$, then $\zg_{\mu\nu}$ is locally a product of two-dimensional
de Sitter space with $S^2$ (if $\Lambda>0$) or two-dimensional anti-de
Sitter space with hyperbolic space (if $\Lambda<0$).
\end{itemize}

\end{itemize}

We conclude that, whatever $\Lambda\in \R$, \emph{static, vacuum, four
dimensional solutions with a degenerate Killing horizon with compact
cross-sections have vanishing rotation one-form $\zh_adx^a$, and
$\mathring A=\Lambda$, with $\zh_{ab}dx^a dx^b$ of constant scalar
curvature $2\Lambda$.}

%In the second
%case, $\zh_a \ne 0$ and the local analysis proceeds as above, with the
%only modification to the conclusion being that in the metric
%\eq{genSchw} we set $U(R) = k - 2M/R - \Lambda R^2/3$.

\bigskip

\noindent{\sc Acknowledgments:} We are grateful to the Isaac Newton
Institute, Cambridge, for hospitality and financial support. HSR is a
Royal Society University Research Fellow.

\bibliographystyle{amsplain}
\bibliography{%
../references/hip_bib,%
../references/reffile,%
../references/newbiblio,%
../references/newbiblio2,%
../references/bibl,%
../references/howard,%
../references/bartnik,%
../references/myGR,%
../references/newbib,%
../references/Energy,%
../references/netbiblio,%
CRT}

\def\cprime{$'$} \def\cprime{$'$}
\providecommand{\bysame}{\leavevmode\hbox to3em{\hrulefill}\thinspace}
\begin{thebibliography}{10}

\bibitem{Ashtekar:2001jb}
A.~Ashtekar, C.~Beetle, and J.~Lewandowski, \emph{Geometry of generic isolated
  horizons}, Class.\ Quantum Grav. \textbf{19} (2002), 1195--1225.

\bibitem{BartnikChrusciel1}
R.~Bartnik and P.T. Chru\'{s}ciel, \emph{{Boundary value problems for
  Dirac-type equations}},  (2003), {math.DG/0307278}.

\bibitem{bunting:masood}
G.~Bunting and A.K.M. Masood{--ul--A}lam, \emph{Nonexistence of multiple black
  holes in asymptotically {E}uclidean static vacuum space-time}, Gen.\ Rel.\
  Grav. \textbf{19} (1987), 147--154.

\bibitem{CarterJMP}
B.~Carter, \emph{Killing horizons and orthogonally transitive groups in
  space--time}, Jour.\ Math.\ Phys. \textbf{10} (1969), 70--81.

\bibitem{Chstatic}
P.T. Chru\'{s}ciel, \emph{The classification of static vacuum space--times
  containing an asymptotically flat spacelike hypersurface with compact
  interior}, Class.\ Quantum Grav. \textbf{16} (1999), 661--687, gr-qc/9809088.

\bibitem{CT}
P.T. Chru\'{s}ciel and K.P. Tod, \emph{The classification of static
  electro-vacuum space--times containing an asymptotically flat spacelike
  hypersurface with compact interior},
  \url{http://www.newton.cam.ac.uk/preprints/NI05067.pdf}.

\bibitem{ChWald}
P.T. Chru\'{s}ciel and R.M. Wald, \emph{On the topology of stationary black
  holes}, Class.\ Quantum Grav. \textbf{11} (1994), L147--152.

\bibitem{GallowaySchoen}
G.J. Galloway and R.~Schoen, \emph{A generalization of {Hawking's} black hole
  topology theorem to higher dimensions},  (2005), Newton Institute preprint
  NI05053-GMR, gr-qc/0509107.

\bibitem{GinspargPerry}
P.H. Ginsparg and M.J. Perry, \emph{Semiclassical perdurance of de {S}itter
  space}, Nucl.\ Phys. \textbf{B222} (1983), 245--268.

\bibitem{LP1}
J.~Lewandowski and T.~Paw\l owski, \emph{Extremal isolated horizons: A local
  uniqueness theorem}, Class.\ Quantum Grav. \textbf{20} (2003), 587--606,
  gr-qc/0208032.

\bibitem{VinceJimcompactCauchyCMP}
V.~Moncrief and J.~Isenberg, \emph{Symmetries of cosmological {C}auchy
  horizons}, Commun. Math. Phys. \textbf{89} (1983), 387--413.

\bibitem{LP2}
T.~Paw\l owski and J.~Lewandowski, \emph{Quasi-local rotating black holes in
  higher dimension: geometry}, Class.\ Quantum Grav. \textbf{22} (2005),
  1573--1598.

\bibitem{LJP}
T.~Paw\l owski, J.~Lewandowski, and J.~Jezierski, \emph{Spacetimes foliated by
  {Killing} horizons}, Class. Quantum Grav. \textbf{21} (2004), 1237--1252.

\bibitem{Reall:2002bh}
H.S. Reall, \emph{Higher dimensional black holes and supersymmetry}, Phys. Rev.
  \textbf{D68} (2003), 024024, hep-th/0211290.

\bibitem{ScottSzekeresII}
S.M. Scott and P.~Szekeres, \emph{The {C}urzon singularity. {II: G}lobal
  picture}, Gen.\ Rel.\ Grav. \textbf{18} (1986), 571--583.

\bibitem{Vishveshwara}
C.V. Vishveshwara, \emph{Generalization of the {``{S}chwarzschild Surface''} to
  arbitrary static and stationary metrics}, Jour.\ Math.\ Phys. \textbf{9}
  (1968), 1319--1322.

\end{thebibliography}
\end{document}